\documentclass{iopart}
\usepackage[dvips]{graphicx}

\newcommand{\be}{\begin{equation}}
\newcommand{\ee}{\end{equation}}
\newcommand{\ba}{\begin{eqnarray}}
\newcommand{\ea}{\end{eqnarray}} 

\newcommand{\heff}{H_{\rm eff}}

\newcommand{\cpt}{\cal{PT}}
\newcommand{\varkappa}{\cal \varsigma}

\begin{document}
\date{\today}
\title{The role of exceptional points in quantum systems}
\author{Ingrid Rotter}
\address{Max-Planck-Institut f\"ur Physik komplexer
Systeme, D-01187 Dresden, Germany }

\begin{abstract}
Exceptional points are known in the mathematical literature for many years. 
They are singular points at which (at least) two eigenvalues of an operator
coalesce. In physics, they can be studied best when the Hamiltonian of the
system is  non-Hermitian. Although the points themselves can not be 
directly identified in physics, their strong influence onto the
neighborhood can be traced. Here, the exceptional points are called
mostly crossing points (of the eigenvalue trajectories) or branch points 
or double poles of the $S$ matrix. 
In the present paper, first the mathematical basic properties of the
exceptional points are discussed. Then, their role in the description 
of real physical quantum systems is considered (after solving the 
corresponding equations exactly). The Hamiltonian of these systems 
is non-Hermitian due to their embedding into an environment (continuum
of scattering wavefunctions). Outside the energy window coupled
directly to the continuum, the Hamiltonian is Hermitian but with
corrections originating from the principal value integral 
of the coupling term via the continuum.
Most interesting value of the non-Hermitian quantum physics 
is the phase rigidity of the eigenfunctions which varies
(as function of a control parameter)
between 1 (for distant non-overlapping states) and 0 (at the
exceptional point where the resonance states completely overlap).
This variation allows the system to incorporate environmentally induced
effects. In the very neighborhood of a crossing (exceptional) point, 
the system can be described well by a conventional nonlinear
Schr\"odinger equation. Here, the entanglement of the different states is
large. In the regime of overlapping resonances, many eigenvalue
trajectories cross or avoid crossing, and
spectroscopic redistribution processes occur in the 
whole system. As a result, a dynamical phase transition takes place
to which all states of the system contribute:   
a few short-lived resonance states are aligned to the scattering states
of the environment by trapping  the other  states. 
The trapped resonance states are long-lived, show chaotic features,
and are described well by means of statistical ensembles.
Due to the alignment of a few states with the states of the environment, 
observable values (e.g. the transmission through the system)
are enhanced. The dynamical phase transition
breaks the spectroscopic relation of the short-lived and long-lived
resonance states to the original individual states of the system. 
These results hold also for $\cpt$ symmetric systems.
The dynamical phase transition characteristic of non-Hermitian quantum
physics, allows us to understand some experimental results 
which remained puzzling in the framework of conventional Hermitian 
quantum physics. The effects caused by the exceptional (crossing)
points in physical systems allow us to manipulate them for 
many different applications.

\end{abstract}


\maketitle

\section{Introduction}
\label{int}

Many years ago, Kato \cite{kato} introduced the notation {\it exceptional
  points} for singularities appearing in the perturbation theory for linear
operators. Consider a family of operators of the form 
\begin{eqnarray}
T({\varkappa} )=T(0)+\varkappa T'
\label{int1}
\end{eqnarray}
where  $\varkappa$ is a scalar parameter, $T(0)$  is the unperturbed operator
and $\varkappa T'$  is the perturbation. Then the
number of eigenvalues of $T(\varkappa )$ is independent of $\varkappa$
with the exception of some special values of $\varkappa$ 
where (at least) two eigenvalues coalesce. These special values of 
$\varkappa$ are the  {\it exceptional points}.
An example is the operator 
\begin{eqnarray}
T(\varkappa) = \left(
\begin{array}{cc}
1 & \varkappa \\
\varkappa & -1 
\end{array} \right) \,.
\label{int2}
 \end{eqnarray}
In this case, the two values $\varkappa = \pm$ i  give the same eigenvalue ~0.

Operators of the type (\ref{int2}) appear in the description of physical
systems, for example in the theory of open quantum
systems \cite{top}. In this case, they represent a  $2\times 2$ Hamiltonian
describing a two-level system with the unperturbed energies 
$\epsilon_1$ and $\epsilon_2$ and the interaction $\omega$ between the two 
levels,
\begin{eqnarray}
H(\omega) = \left(
\begin{array}{cc}
\epsilon_1 & \omega \\
\omega & \epsilon_2 
\end{array} \right) \, .
\label{int3}
 \end{eqnarray}
In an open quantum system, two states  
can interact directly (corresponding to a first-order term)  
as well as via an environment (second-order term) \cite{top}.  
In the present paper, we consider the
case that  the direct interaction is contained  in the
energies $\epsilon_{k}~(k=1,2)$. Then $\omega$ contains  exclusively
the coupling of the states via the environment which, 
in the case of an open quantum system, consists 
of the continuum of scattering wavefunctions into which the
system is embedded. 
This allows to study  environmentally induced effects in 
open quantum systems in a very clear manner \cite{top}.

The eigenvalues of the operator $H(\omega)$ are
\begin{eqnarray}
\varepsilon_{1,2}&=&\frac{\epsilon_1 + \epsilon_2}{2} \pm Z \; ; \; \quad
Z=\frac{1}{2}\sqrt{(\epsilon_1 - \epsilon_2)^2 + 4 \omega^2} \; .
\label{int4}
\end{eqnarray}
The two eigenvalue trajectories cross when $Z=0$, i.e. when
\begin{eqnarray}
\frac{\epsilon_1 - \epsilon_2}{2\omega} = \pm \, i \; .
\label{int5}
\end{eqnarray}
At these {\it crossing points}, the two eigenvalues coalesce,
\begin{eqnarray}
\varepsilon_1 ~=~ \varepsilon_2 ~\equiv ~\varepsilon_0 \, . 
\label{int6}
\end{eqnarray}
The crossing points may be called therefore exceptional points.

However, there are some essential differences between the
exceptional points considered in the mathematical literature and the
crossing points which appear in physical systems. The differences arise
from the fact that the crossing points are points in the continuum of
scattering wavefunctions (which represents the environment). They are 
therefore of measure 0 and can not be observed directly. However, they
influence the behavior of the eigenvalue trajectories $\varepsilon _k
(\alpha)$ (where $\alpha$ is a certain parameter) in their
neighborhood  in a non-negligible manner. Thus, the most interesting
features of the exceptional (crossing) points in physical systems are
not the properties at the crossing points  themselves.  Much more
interesting are
their effects onto the eigenvalue trajectories $\varepsilon_k
(\alpha)$ in a finite  parameter range around the critical value
$\alpha  = \alpha_{\rm cr}$ (at which two trajectories cross) 
and, above all, the behavior of the   
eigenvalue trajectories in approaching the crossing point,
$\varepsilon_k(\alpha) \to \varepsilon_k (\alpha_{\rm cr})$. 
The phenomenon of  {\it avoided level
crossing} is known in physical systems since many years \cite{landau}.
It occurs not only for discrete states but also for narrow resonance states
\cite{ro01}.
In the scattering theory, the crossing points cause {\it double
poles of the $S$ matrix}. For details see \cite{top}. 
 
According to their influence on many physical observables, 
exceptional points are considered under different aspects in the 
physical literature.   The topological 
features are theoretically considered by means of a $2\times 2$ system
in, e.g., \cite{mondragon,heiss,hemuro} 
and experimentally studied on a microwave cavity in
\cite{demb1,demb2}. The results of recent theoretical studies can 
be found in \cite{ali,holo}. A topological
transition in a non-Hermitian quantum walk is discussed in \cite{rudner}.

In many-level quantum systems, the exceptional points are called
crossing points of eigenvalue trajectories,
e.g. \cite{frwi,solov,top}, or double poles of the $S$ matrix,
e.g. \cite{ro91,marost2,marost3,plo}, or branch points, 
e.g. \cite{double}. In most studies, the biorthogonality 
of the eigenfunctions of the Hamiltonian plays an important role
and is considered explicitely.
In \cite{marost2,marost3}, laser-induced continuum structures in atoms
are studied. They are of interest especially in the neighborhood 
of crossing points (double poles of the $S$ matrix). The high-order 
harmonic generation in a driven two-level
atom is related to abrupt population transfers between states at the
avoided level crossings \cite{carla}. Recently, the laser control of 
vibrational transfer based on exceptional points is studied in 
\cite{atabek}. The influence of exceptional points on the photoionization
cross section is  investigated  in \cite{cart1}. 
Also in nuclear physics, exceptional points at low energy 
appear for realistic values of the coupling to the continuum  \cite{plo}. 
The relation between exceptional points and the Petermann factor
characterizing the enhancement in intrinsic laser line widths and spontaneous
emission rates, is discussed in \cite{peterm}. In \cite{top}, the phase
rigidity of the eigenfunctions of the non-Hermitian Hamilton operator in
approaching an exceptional point is related to environmentally induced effects
in quantum systems. This relation holds also for $\cpt$ symmetric 
systems \cite{jopt}. The phase rigidity is shown to be anti-correlated 
with the transmission probability through quantum dots \cite{burosa}. 

The relation between exceptional points on the one hand, and the phenomenon 
of resonance trapping and dynamical phase transitions, on the other hand,
is studied theoretically in many papers, 
see the recent review \cite{top}. It is proven experimentally       
in \cite{stm}. As a result of resonance trapping in ${\cal P}$ symmetric
systems, bound states in the continuum may appear \cite{marost2,rszero}. The
relation between avoided level crossings and bound states in the continuum
(the last phenomenon is called mostly population trapping in atomic physics), 
is first obtained in \cite{frwi}. 
In \cite{shot}, the statistical properties of the trapped states are
considered. In \cite{schom}, the 
lifetimes of electromagnetic quasibound states in dielectric 
microresonators with fully chaotic ray dynamics are 
statistically analyzed. The necessary 
renormalization  is linked to the formation of short-lived resonances, 
i.e. to the resonance trapping mechanism. According to \cite{jmp}, the 
resonance states of a many-body system at high level density are
described well by a statistical ensemble containing the interaction
between {\it all} particles (e.g. the Gaussian orthogonal ensemble), 
while those at low level density are described best by a combination
of one-body problems (e.g. the shell model).
Meanwhile, an exceptional point is observed directly 
in a chaotic optical microcavity \cite{korea}.
The influence of exceptional points 
in quantum chemistry is studied some time ago \cite{miller}. 
Recently, quantum dynamical phase transitions are found
experimentally and theoretically in the spin swapping operation 
\cite{pastawski,swap,swap2}. 

Resonance coalescence in molecular photodissociation is studied in
\cite{mois}. 
The visualization of an exceptional point in a $\cpt$ symmetric 
directional coupler is demonstrated in \cite{moisgue}.  
In \cite{peskin} it is shown that the nature of the transport through a
molecular junction is determined by a dimensionless parameter which
measures the degree of resonance overlap in the system. 
Experimental studies in quantum point contacts show the importance of
detector backaction \cite{bird}. In a $\cpt$-symmetric square well,
bound states appear below a certain threshold of the degree of 
non-Hermiticity, while beyond the threshold the two lowest real
energies are shown analytically to merge and disappear \cite{znojil}. 
The  phase lapses observed
experimentally \cite{heiblum1,heiblum2} in the transmission through small 
quantum dots, can be explained qualitatively \cite{murophas} 
by the dynamical phase transition occurring in the regime of overlapping
resonances. 

Exceptional points are found to play a role also in Bose-Einstein
condensates of gases \cite{cart2}. In the quantum motion of a
Bose-Einstein condensate in an optical cavity, the Dicke-model phase
transition is observed \cite{nagy}, what is nothing but the resonance 
trapping phenomenon \cite{top}.  The doorway states in nuclear
reactions can be considered to be a manifestation of the Dicke model 
super-radiant mechanism \cite{soze,auerbach}.
The appearance of exceptional points is studied recently 
even in classical systems: in cosmic 
structure formation, where the magnetorotational instability 
is known to play an important role \cite{kirillov}. Here,
the mechanism of instability transfer between modes 
through a spectral exceptional point is identified, which allows to
explain some data. 

The aim of the present paper is to give a consistent 
and unifying representation of the
role of singular (exceptional) points in quantum systems. The
constraints  originating from the physical 
boundary conditions are taken into account. Among others, it will be shown
that exceptional points influence not only the resonance states,
but also the discrete states of the system. Here, they cause the avoided level
crossing phenomenon known since many years \cite{landau}, as well as effective
forces used in almost all numerical calculations.    
Most interesting is the regime of overlapping resonances where the meaning of 
exceptional points for physical processes 
and their impact on the dynamics of the system can be controlled.
Here, symmetry breaking caused by exceptional points plays an
important role.

The paper is organized in the following manner. In sections \ref{eiv}
and \ref{eif}, the eigenvalues and eigenfunctions of a $2\times 2$ Hamilton
operator of the type (\ref{int3}) are considered. Here, the basic properties
of the exceptional (crossing) points are sketched. At these singular points,
level repulsion passes into width bifurcation. The eigenfunctions 
$\phi_k$ of the non-Hermitian Hamilton operator $H$ are biorthogonal leading
to some freedom for their normalization (since 
$\langle \phi_k^*|\phi_l\rangle $ is not necessarily a real number).
We choose  $\langle \phi_k^*|\phi_l\rangle = \delta_{kl}$ 
in order to describe
the transition from overlapping to non-overlapping resonance states
in a smooth manner (the last ones are normalized 
as $\langle \phi_k|\phi_l\rangle = \delta_{kl}$  according to
conventional quantum theory). As a consequence, the phases of the
eigenfunctions of $H$ are not rigid in approaching an exceptional (crossing)
point. This mathematical result is surely the most interesting one of
non-Hermitian quantum physics. It allows the system to incorporate 
environmentally induced effects (feedback from the coupling to the 
environment).  In the neighborhood of the crossing points, the system is
described well by a conventional nonlinear Schr\"odinger equation.

Section \ref{man} shows that the basic results of the $2\times 2$ problem 
survive when realistic systems with many levels are considered.
The eigenvalues of the Hamiltonian
are complex or real, according to the boundary conditions. In the
first case, the eigenstates are resonant (with, usually, a finite lifetime) 
while they are discrete (corresponding to an infinitely long lifetime) 
in the second case.  The coupling of the states via the continuum
becomes important in the regime of overlapping resonances. For the 
discrete states, it introduces effective forces. 
Section \ref{sol} gives the solution $\Psi_c^E$ of the Schr\"odinger
equation in the total function space, including discrete and scattering states.
The Hamilton operator of the whole system is Hermitian. The solution $\Psi_c^E$ 
is found by using a projection operator formalism. The two subspaces 
correspond to  {\it system} (localized in space) and {\it environment}
(extended in space). The solution $\Psi_{c ~int}^E $
inside the localized part of the system can be represented in a set of
biorthogonal wavefunctions. Hence, the phases of the $\Psi_{c ~int}^E $
are not rigid such that an alignment of some of them with the channel wave
functions of the environment is possible also in the many-level case. 
This alignment occurs by trapping other resonance states, i.e. 
by width bifurcation. In section \ref{sma}, the $S$ matrix is given
by using the $\Psi^E_c$. Most interesting are the double poles 
of the $S$ matrix appearing at the crossing points. Here, 
the line shape of the resonance shows  nonlinear effects.

The entanglement of the states is considered in section \ref{mix}.
It is most interesting in the regime of overlapping resonances where 
many true and avoided level crossings occur. In section
\ref{avd}, the avoided level crossing phenomenon is traced, by means of a
control parameter,  from resonance 
states in the overlapping regime to narrow resonance states and finally to
discrete states. The relation to quantum chaos is discussed.
In section \ref{dyn}, the
interplay between system and environment is discussed. It is shown that the
entanglement of the states via the continuum 
occurring in the regime of overlapping resonances of a many-level
system,  is nothing but a dynamical phase transition. 
At and in the neighborhood of the crossing points, 
the resonance states lose their individual spectroscopic features
under the influence of the environment. The aligned states cause some
{\it transparency} of the system while the trapped states are 
described best by a statistical ensemble, e.g. 
by the Gaussian orthogonal ensemble. 
In section \ref{exp}, some experimental results are sketched which are
puzzling in conventional Hermitian quantum physics, but may be 
explained (at least qualitatively) by means of dynamical phase
transitions, i.e. by considering the exceptional points characteristic of
non-Hermitian quantum physics. The results are summarized in the last
section.

\section{The eigenvalues of a non-Hermitian $2\times 2$ 
Hamilton operator}
\label{eiv}

We consider the  Hamiltonian (\ref{int3}) 
with the unperturbed energies $\epsilon_i ~(i=1,2)$ of the two states
and the interaction $\omega$ between them. The interaction $\omega$
contains exclusively the coupling of the states
via the {\it environment}, which consists of the continuum of decay
channels into which the states are embedded. 
The interaction $\omega$ is therefore a second-order interaction term.    
The two eigenvalues $\varepsilon_{k} ~(k=1,2)$ of  (\ref{int3}) 
are given in (\ref{int4}). 
The Hamiltonian may be Hermitian or non-Hermitian.

For a Hermitian operator, the unperturbed
energies $\epsilon_i $ of the states are real.
The interaction $\omega$ being  the principal
value integral of the coupling term via the continuum, is also real 
\cite{top}. Accordingly, the two eigenvalue trajectories 
$\varepsilon_i(\alpha)  = e_i(\alpha)$ (where $e_i(\alpha)$ is real)
cannot cross (for $\omega \ne 0$) when traced as a function of a 
certain parameter $\alpha$, see (\ref{int4}).
Instead, they  avoid crossing. This
phenomenon is very well known for about 70  years \cite{landau}. 
The fictive crossing point is called 
{\it diabolic point}. The topological structure of this point is
characterized by the Berry phase \cite{berry}
which is studied theoretically and experimentally in many  papers.

The situation is another one for a non-Hermitian operator.  In such a
case, the unperturbed energies $\epsilon_i$ are usually complex. Also the
interaction $\omega$ is complex, in general, since it contains the
principal value integral as well as the residuum of the coupling term 
describing the interaction of the two states via the environment
(continuum of scattering wavefunctions) \cite{top}.
The states  can decay, in general, and
the two eigenvalues of (\ref{int3}) can be written as
\begin{eqnarray} 
\varepsilon_{1,2}= 
e_{1,2} - \frac{i}{2} ~\gamma_{1,2}  \qquad ({\rm with} ~\gamma_{1,2}
\ge 0 ) \; .
\label{eiv1}
\end{eqnarray}
The widths $\gamma_i$ are proportional to the inverse lifetimes
$\tau_i^{-1}$ of the  states, $i=1,2$.
The two eigenvalue trajectories $\varepsilon_i(\alpha)$
may cross according to (\ref{int4}) and (\ref{int6}). The crossing point 
is an exceptional point in agreement with the definition given in
\cite{kato}, see (\ref{int1}) and (\ref{int2}). 
The topological phase of the exceptional point is  twice the Berry
phase \cite{top,hemuro}. This theoretical result is proven experimentally
by means of a microwave cavity \cite{demb1}.

According to the eigenvalue equation (\ref{int4}) $Z$ is  complex,
usually. Re$(Z)$ causes repulsion of the levels in energy. 
This result corresponds to the avoided level crossing phenomenon 
known for discrete states since many years \cite{landau}. 
It is the dominant part also in the case when the resonance states are narrow
(long-lived), i.e. when the interaction $|\omega|$ of the states 
via the continuum of scattering wavefunctions is small.
The value Im$(Z)$ has another physical meaning. It is the dominant
part when $|\omega|$ is large what is the case, above all, when
the two resonances overlap. According to (\ref{int4}), 
Im$(Z)$ is related to a bifurcation of the widths of the levels.  

Due to width bifurcation, resonance states with  long
lifetime may appear together with short-lived states. The time scales
characterizing  these two different types of states, 
may differ strongly from one another.
It is possible even that the widths of some states vanish, i.e.
that $\gamma_i =0$ for some states. These states with vanishing width 
are called, usually, {\it bound states in the
continuum} \cite{frwi}.  Examples are studied in calculations for 
laser-induced continuum structures in atoms \cite{marost2}  
as well as for the transmission through quantum dots \cite{rszero}. 
In these calculations, resonance states with zero width appear
at realistic parameter values. Tracing their appearance as
a function of a parameter, one can see that they are nothing but
special resonance states. The only hint in the cross section
to such a state is the (elastic) scattering  phase shift which passes 
into a  jump by $\pi$ at the 
energy of the state, see Fig. 5 in \cite{rszero} for an example.
These bound states in the continuum coexist  with short-lived states. 
In \cite{alispec,longhi}, the bound states in the continuum are called
spectral singularities.  

Hence, the real and imaginary parts of the
complex eigenvalues (\ref{eiv1}) of the Hamiltonian (\ref{int3}) 
have a physical meaning in a quantum system in which the localized 
states of the system are embedded in an extended continuum of
scattering states. The real parts $e_i$ stand for the positions
in energy of the (almost) localized states while the imaginary parts 
$\gamma_i$ give the  widths (inverse lifetimes) of these states. 
It is  $\gamma_i >0$ when the decay of the states is not
forbidden by any selection rule (and when the states are inside the energy
window coupled to the continuum). The decay is an irreversible process
\cite{top}. Only at strong coupling to the continuum  [corresponding to
Im$(Z) \gg$ Re$(Z)$ in (\ref{int4})], discrete states may appear
due to width bifurcation also inside the energy window coupled to the
continuum.

Starting with the papers \cite{bender,benderrev} by Bender et al., 
it has been  shown that 
a wide class of ${\cal PT}$ symmetric non-Hermitian Hamiltonians 
provides entirely real spectra. In order to realize 
complex $\cpt$ symmetric structures, the formal equivalence
of the quantum mechanical Schr\"odinger equation to the optical wave
equation in $\cpt$ symmetric optical lattices can be exploited
by involving symmetric index guiding and an antisymmetric gain/loss profile 
\cite{ruschhaupt,makris1,makris2,makris3}.
Meanwhile,  experimental studies are performed. The results 
given in \cite{guo}  have confirmed  the expectations and have,
furthermore, demonstrated the onset of passive $\cpt$ symmetry
breaking within the context of optics. This phase transition was found
to lead to a loss induced optical transparency in specially designed
pseudo-Hermitian potentials. In \cite{nature}, the wave propagation in
an active $\cpt$ symmetric coupled wave guide system is studied. 
Both spontaneous $\cpt$ symmetry breaking and power oscillations
violating left-right symmetry are observed. Moreover, the relation  
of the relative phases of the eigenstates of the system to their 
distance from the level crossing (exceptional) point is obtained. 
Approaching this point, the phase transition occurs. 
In \cite{makris4}, the Floquet-Bloch modes in $\cpt$ symmetric optical
lattices are examined in detail.

Thus, the formal equivalence of the optical wave
equation in $\cpt$ symmetric optical lattices to 
the quantum mechanical Schr\"odinger equation allows us to study the 
properties of quantum systems the states of which can not only decay 
due to their coupling to the environment, but may also be formed
out of the environment due to this coupling. In optics, these two 
possibilities are called {\it loss} and {\it gain}. The theory  
contains both  possibilities. This fact makes the
study of $\cpt$ symmetric optical lattices  a very attractive one.

In   $\cpt$ symmetric optical lattices, the eigenvalues are
\begin{eqnarray}
\varepsilon_{1,2} = 
e_{1,2} \pm \frac{i}{2} ~\gamma_{1,2}  \qquad ({\rm with} 
~\gamma_{1,2} \ge 0 \; {\rm ~and} \; ~e_{1} = e_2) 
\label{eiv2}
\end{eqnarray}
in difference to (\ref{eiv1}). Due to $\cpt$ symmetry, all 
eigenvalues $\varepsilon_i = e_i$ may be real 
(corresponding to $\gamma_i =0$) 
when Re$(Z) \gg$ Im$(Z)$ in (\ref{int4}), i.e. at low
coupling of the states to the continuum. Under these conditions,
the optical wave equation describes a reversible process.
However, the $\cpt$ symmetry breaks  at  Im$(Z) \gg$ Re$(Z)$
and then $\gamma_{1,2} \ne 0$. 

It follows immediately that the $\cpt$ symmetric models can not be
mapped onto models of open quantum systems \cite{jopt}, although
formally such a mapping seems to be possible by adding a
constant imaginary energy shift to the eigenvalues. Both models differ
fundamentally from one another when applied to the description of
physical systems. It is this difference between the two models which
will allow us to receive interesting information on quantum systems by
studying not only open quantum systems (which exist in nature) but
also  $\cpt$ symmetric systems (which are formally
equivalent to them).

\section{The eigenfunctions of a non-Hermitian $2\times 2$
Hamilton operator}
\label{eif}

The eigenfunctions of the non-Hermitian Hamilton operator $H$ are
biorthogonal,
\begin{eqnarray}
\langle \phi_k^*|\phi_l\rangle  = \delta_{k, l} 
\; .
\label{eif1}
\end{eqnarray}
From these equations follows
\begin{eqnarray}
\langle \phi_k|\phi_{k }\rangle & \equiv &  A_k \ge 1 
\label{eif2}
\end{eqnarray} 
and 
\begin{eqnarray}
\langle \phi_k|\phi_{l\ne k}\rangle =- 
\langle \phi_{l \ne k  }|\phi_k\rangle & \equiv & B_k^l ~; 
~~~|B_k^l|\ge 0 \; .
\label{eif3}
\end{eqnarray}
At the crossing point 
\begin{eqnarray}
A_k^{\rm (cr)} \to \infty \qquad \quad 
|B_k^{l ~{\rm (cr)}}| \to \infty \; ,
\label{eif4}
\end{eqnarray}
for details see \cite{top}. 

The relation between the eigenfunctions 
$\phi_1$ and $\phi_2$ of the operator (\ref{int3}) at the crossing 
point  is
\begin{eqnarray}
\phi_1^{\rm cr} \to ~\pm ~i~\phi_2^{\rm cr} \quad \qquad \phi_2^{\rm cr} \to
~\mp ~i~\phi_1^{\rm cr}  
\label{eif5}
\end{eqnarray}  
according to analytical  \cite{ro01,gurosa} as well as numerical
studies \cite{marost3}.
The two eigenfunctions are linearly dependent of one another at the
crossing point such that the number of eigenfunctions of $H$
is reduced at this point. 
This result shows once more that the crossing point is an exceptional
point in the sense defined by Kato \cite{kato}.

In an experimental study on a microwave cavity \cite{demb1}, the topological
structure of the exceptional point and its surrounding is studied by
encircling it and tracing the relative amplitudes of the 
wavefunctions (field distributions inside the cavity). As a result,  
the wavefunctions including their phases are restored after four
surroundings. The authors \cite{demb1} interpreted the experimental
data by two theoretical assumptions: (i)
the two wavefunctions coalesce into one at the exceptional point,
$\phi_1^{\rm cr} \leftrightarrow \phi_2^{\rm cr} $, and (ii)
only one of the wavefunctions picks up a phase of $\pi$ (a sign change) 
when encircling the  critical point. With these two assumptions,  
the wavefunctions are restored after four surroundings as found
experimentally.   

The experimental result  can be explained, however, without any
additional assumptions by using the relations (\ref{eif5})
\begin{eqnarray}
\label{cycle}
{\rm 1. ~~cycle:\, } \qquad \varepsilon_{1,2} \to \varepsilon_{2,1}    \qquad 
~~~~~~~\phi_{1,2} &\to& ~\pm\,i\,\phi_{2,1} \nonumber \\
{\rm  2. ~~cycle: } \qquad \varepsilon_{2,1} \to \varepsilon_{1,2}    \qquad 
~~\pm\, i\,\phi_{2,1} &\to& -\phi_{1,2} \nonumber \\
{\rm \, 3. ~~cycle:} \qquad \varepsilon_{1,2} \to \varepsilon_{2,1}    \qquad 
~~~~-\phi_{1,2} &\to& \mp \, i\,\phi_{2,1} \nonumber \\
{\rm 4. ~~cycle:\, } \qquad \varepsilon_{2,1} \to \varepsilon_{1,2}    \qquad 
~~\mp \, i  \,\phi_{2,1} &\to& ~~~\phi_{1,2} 
\label{eif6}
\end{eqnarray}
As can be seen, the eigenvalues are restored after two surroundings
and  the eigenfunctions are restored  after four
surroundings,  in full agreement with the experimental result.

In any case,
$|~\phi_1^{\rm cr}~|~ = ~|~\phi_2^{\rm cr}~| $
at the crossing point in agreement with the statement that the 
number of eigenstates is reduced at the exceptional point.
The topological phase is twice the Berry phase, in accordance with the
enlarged function space in open quantum systems.

Theoretical studies \cite{gurosa} have shown that 
{\it associated vectors} $\phi_i^{cra}$ defined by the Jordan relations, 
appear at the crossing points. The corresponding equations are
\begin{eqnarray}
(H -\varepsilon_0)~\phi_{1,2}^{\rm cr} &=& 0 
\nonumber \\
(H -\varepsilon_0)~\phi_{1,2}^{\rm cra} &=& \phi_{1,2}^{\rm cr} \; .  
\label{eif8}
\end{eqnarray}
The existence of two states in the very neighborhood of the 
exceptional point has
been seen in a numerical calculation for the elastic scattering of a
proton on a light nucleus \cite{plo}. The elastic scattering phase
shifts  jump always by
$2\pi$ (and not by $\pi$ as for a single resonance state).

Furthermore, the phases of the wavefunctions jump by $\pi /4$ 
at the crossing point (when traced as a function of a parameter) 
due to the biorthogonality  (\ref{eif1})
of the eigenfunctions of the non-Hermitian Hamiltonian $H$,
see also (\ref{eif3}). This result has been proven in many numerical 
studies, see \cite{top}.

Let us now consider the consequences of the biorthogonality relations
(\ref{eif1}) and (\ref{eif2})
for the two borderline cases characteristic of neighboring resonance states.  
\begin{enumerate}
\item
 The two levels are distant from one another. Then the eigenfunctions
 are (almost) orthogonal 
\begin{eqnarray}
\langle \phi_k^* | \phi_k \rangle   \approx
\langle \phi_k | \phi_k \rangle  = A_k \approx 1 \; .
\label{eif9}
\end{eqnarray}
\item
The two levels cross. Then the two eigenfunctions are linearly
dependent according to (\ref{eif5}) and 
\begin{eqnarray}
\langle \phi_k | \phi_k \rangle = A_k \to \infty \; .
\label{eif10} 
\end{eqnarray}
according (\ref{eif4}).
\end{enumerate}
The two relations (\ref{eif9}) and (\ref{eif10}) 
show that the phases of the two eigenfunctions
relative to one another change when the crossing point is approached. 
This can be expressed quantitatively by defining the {\it phase
  rigidity} $r_k$ of the eigenfunctions $\phi_k$,
\begin{eqnarray}
r_k \equiv \frac{\langle \phi_k^* | \phi_k \rangle}{\langle \phi_k 
| \phi_k \rangle} = A_k^{-1} \; . 
\label{eif11}
\end{eqnarray}
According to (\ref{eif9}) and (\ref{eif10}) holds 
\begin{eqnarray}
1 ~\ge ~r_k ~\ge ~0 \; .  
\label{eif12}
\end{eqnarray}
The  non-rigidity $r_k$ of the phases of the eigenfunctions of $H$ 
follows also from the fact that $\langle\phi_k^*|\phi_k\rangle$
is a complex number (in difference to the norm
$\langle\phi_k|\phi_k\rangle$ which is a real number) 
such that the normalization condition
(\ref{eif1}) can be fulfilled only by the additional postulation 
Im$\langle\phi_k^*|\phi_k\rangle =0$ (what corresponds to a rotation
\cite{top}).

The variation of $r_k$ according to (\ref{eif12}) 
in approaching the crossing point of
two eigenvalue trajectories is proven experimentally by means of a
study on a microwave cavity \cite{demb2}. As a result of the
experimental study, 
the phase difference between two modes is $\pi$ at large distance and
decreases to $\pi /2$ at the crossing point.

The authors of \cite{demb2} interpreted the experimental data
by assuming (i) that the singular point is  a chiral state 
(in spite of the phase jump occurring at the crossing point,
when traced as a function of a certain parameter),
(ii) that the number of states is reduced from 2 to 1 at the crossing
point (in spite of the existence of the associate vector (\ref{eif8}))
and (iii) that a single point in the continuum can be identified
(although it is of measure zero).
The authors are unable to explain  the large parameter range 
in which the phase difference decreases in approaching the crossing point.

Considering the phase rigidity $r_k$ in the regime of the two
overlapping resonance states, no  additional assumptions are required 
for the explanation of the experimental results given in \cite{demb2},
since the phase rigidity (being a quantitative measure for the degree 
of resonance overlapping) varies smoothly in a comparably large 
parameter range. It can therefore be concluded that the experimental
results  \cite{demb2}
prove the statement that the phases of the eigenfunctions of the
non-Hermitian Hamilton operator $H$ are not rigid in approaching the 
crossing point, but vary according to (\ref{eif9}) to (\ref{eif12}). 

According to (\ref{int3}), the
Schr\"odinger equation with the unperturbed Hamilton operator 
$H_0$ and a source term
arising from the interaction $\omega$ with another state 
via the continuum of scattering states reads \cite{ro01}
\begin{eqnarray}
\label{mix1}
(H_0  - \epsilon_n) ~| \phi_n \rangle & = &
- \left(
\begin{array}{cc}
0 & \omega \\
\omega & 0 
\end{array} \right) |\phi_n \rangle 
\equiv  W ~| \phi_n \rangle \nonumber \\
&=& \sum_{k=1,2} \langle \phi_k |W|\phi_n\rangle \sum_{l=1,2}
\langle \phi_k|\phi_l\rangle |\phi_l\rangle \nonumber \\
&=& \sum_{k=1,2} \langle \phi_k |W|\phi_n\rangle
\{ A_k ~|\phi_k\rangle + 
\sum_{l\ne k} ~B_k^l ~|\phi_l\rangle \} \; .
\end{eqnarray}
Here
$\langle \phi_k|\phi_{k }\rangle  \equiv   A_k  \ge 1 $ according to 
(\ref{eif2}) and $
\langle \phi_k|\phi_{l\ne k }\rangle = - 
\langle \phi_{l \ne k  }|\phi_{k}\rangle  \equiv 
B_k^{l}, ~|B_k^{l}|\ge 0 
$ according to  (\ref{eif3}). The $A_k$ and $B_k^l$
characterize the degree of resonance overlapping.
In the regime of overlapping resonances, $1>A_k >0$, $|B_k^l| >0$, and 
equation (\ref{mix1}) is nonlinear. 
The most important part of the nonlinear contributions is contained in 
\begin{eqnarray}
\label{mix2}
(H_0  - \epsilon_n) ~| \phi_n \rangle =
\langle \phi_n|W|\phi_n\rangle ~|\phi_n|^2 ~|\phi_n\rangle  
\end{eqnarray}
which is a nonlinear Schr\"odinger equation. According to
(\ref{mix1}), the nonlinear Schr\"odinger equation (\ref{mix2})
goes over smoothly into a linear Schr\"odinger
equation when departing from the exceptional point, i.e. 
in its neighborhood.

\section{The many-level system}
\label{man}

We consider now a conventional quantum system with $N$ discrete states.
The wavefunctions $\Phi_k^B$ of the states of this system
are eigenfunctions of a Hermitian Hamilton operator $H_B = H_B^0+V$
which is assumed to contain
the direct interaction $V$ between the different states. 
Such a system is localized in space. We assume that this system is,
in a certain energy window, embedded into the extended 
continuum of scattering  wavefunctions $\xi^E_c$.
The energy window is defined by the two threshold energies 
$E_{\rm thr}^l$ and $E_{\rm thr}^h$ determining the conductance band
of, e.g., a quantum dot.  In nuclei, $E_{\rm thr}^h \to \infty$.
In this manner, an {\it open quantum system} is defined. 
In the following, we use this definition.

The mathematical description of this system meets the problem that 
the two  wavefunctions $\Phi_k^B$ and $\xi_c^E$  
are of different type. They are normalized differently,
\begin{eqnarray}
\langle \Phi_i^B |\Phi_j^B \rangle = \delta_{i,j}
\label{man1}
\end{eqnarray}
and 
\begin{eqnarray}
\langle \xi_c^E |\xi_{c'}^{E'} \rangle = \delta(E-E') ~\delta_{cc'} 
\label{man2}
\end{eqnarray}
where $c$ stands for a certain decay channel and $E$ is the energy of
the system. For the channel wavefunctions, the shortened notation
$\xi_c^E$ is used here (see \cite{top}). Also the boundary conditions
are different for the two types of wavefunctions. 

From the mathematical point of view, this problem can be overcome best in
the following manner \cite{feshbach}. It is convenient to 
separate the total
 function space into two subspaces, one of which 
(the  $Q$ subspace) contains the $\Phi_k^B$ 
while the other one (the $P$ subspace) consists of the $\xi^E_c$.
In the two subspaces, the corresponding Schr\"odinger equation 
(including the boundary conditions) can be
solved by using the well-known standard methods. 
With $P+Q=1$, the two subspaces (system and environment) 
are well defined. The solution in the total function space can then be
obtained by combining the solutions obtained in the two subspaces, 
see section \ref{sol}. 

In the energy window coupled directly to the
continuum of scattering wavefunctions, the
discrete states (with infinite lifetime) of the $Q$ subspace 
pass into resonance states (with finite lifetime) due to their embedding 
into the $P$ subspace.
Beyond the energy window, the discrete states remain discrete. 
Thus, also the boundary conditions between the two subspaces
play an important role in considering the many-level system.  

In the open quantum system, the
states of the $Q$ subspace can interact via the common environment,
i.e. via the states of the $P$ subspace. Hence, the 
Hamilton operator consists of a first-order and a second-order 
interaction term,
\begin{eqnarray}
\label{man3}
{H}_{\rm eff}& = & H_{B} + V_{BC} ~\frac{1}{E^+ - H_C}  ~V_{CB}
\end{eqnarray}
with
\begin{eqnarray}
\label{man4}
{\rm Re}\; \{\langle \Phi_i^{B} | {H}_{\rm eff} |  \Phi_j^{B} \rangle \}
& = & \langle \Phi_i^{B} | H_{B} |  \Phi_j^{B} \rangle +
\frac{1}{2\pi} \sum_c {\cal P} 
\int\limits_{E_{\rm thr}^l}^{E_{\rm thr}^h} {\rm d} E' \;  
\frac{\hat\gamma_i^c \hat\gamma_j^c}{E-E'} 
\\ 
\label{man5}
{\rm Im}\; \{\langle \Phi_i^{B} | {H}_{\rm eff} |
  \Phi_j^{B} \rangle\}
& = & 
- \frac{1}{2}\; \sum_{c}  \hat\gamma_i^c \hat\gamma_j^c \, .
\end{eqnarray}
Here, ${\cal P}$ denotes the principal value integral and
\begin{eqnarray}
\label{man6}
\hat \gamma_k^c =  
\sqrt{2\pi}\, \langle \xi^{E}_{c} |V| \Phi_{k}^B \rangle 
\end{eqnarray}
is the coupling matrix element between the wavefunctions of the two
subspaces. The direct (first-order) interaction $V$ is included in
$H_B$ and its eigenfunctions $\Phi_k^B$.

In conventional quantum mechanics, the effective Hamilton operator
$H_{\rm eff}$ is assumed to be Hermitian, i.e. the matrix elements
${\rm Re}
\{\langle \Phi_i^{B} | {H}_{\rm eff} |  \Phi_j^{B} \rangle \} $
are considered to correspond to  effective forces.
The non-Hermitian part is not all considered, i.e. 
${\rm Im}\; \{\langle \Phi_i^{B} | {H}_{\rm eff} |
  \Phi_j^{B} \rangle\} =0 $
is assumed.

Here, we are looking for the {\it exact solution} of the problem. 
We calculate not
only Im~$\{\langle \Phi_i^{B} | {H}_{\rm eff} | \Phi_j^{B} \rangle\}$,
but also Re~$\{\langle \Phi_i^{B} | {H}_{\rm eff} | \Phi_j^{B}
\rangle\}$, and that by including the principal value integral
and without any statistical assumptions. The Schr\"odinger equation reads
\begin{eqnarray}
\label{man10}
(H_{\rm eff} -z_k)\, \Phi_k = 0
\end{eqnarray}
with the eigenvalues $z_k$ and eigenfunctions $\Phi_k$ of $\heff$.
In detail:
\begin{enumerate}
\item
The states inside the energy window are 
coupled directly to the environment such that 
the effective Hamilton operator $\heff$ is non-Hermitian, i.e.
the principal value integral in (\ref{man4}) 
as well as the residuum (\ref{man5}) have to be calculated.
The eigenvalues are complex,
\begin{eqnarray}
z_k = E_k - \frac{i}{2} \, \Gamma_k
\label{man7}
\end{eqnarray}
in general, and the eigenfunctions $\Phi_k $ are complex and biorthogonal,
\begin{eqnarray}
\langle \Phi^*_i|\Phi_j\rangle = \delta_{i,j},
\label{man8}
\end{eqnarray}
compare (\ref{eif1}). The coupling matrix elements between 
the $\Phi_k$ and the $\xi_c^E$ are 
\begin{eqnarray}
\gamma_k^c =  
\sqrt{2\pi}\, \langle \xi^{E}_{c} |V| \Phi_{k} \rangle 
\label{sol8}
\end{eqnarray}
in analogy to (\ref{man6}).
\item
The states outside the energy window are not coupled directly 
to the environment
such that the effective Hamiltonian $\heff$ is Hermitian
at the energy of the states, i.e. 
only the principal value integral in (\ref{man4}) has to be calculated.
At the energy of the states, the
eigenvalues $z_k=E_k$  are real, i.e. $ \Gamma_k = 0$, and
the $\Phi_k $ are orthogonal in the standard manner, 
\begin{eqnarray}
\langle \Phi_i | \Phi_j \rangle = \delta_{i,j}  \; .
\label{man9} 
\end{eqnarray}
The coupling matrix elements (\ref{sol8}) between the 
$\Phi_k$ and the $\xi_c^E$ vanish at the energy of the state.
They are, however, different from zero at energies inside the window coupled
directly to the environment and contribute to the principle value integral.
\end{enumerate}
Thus, the non-Hermitian Hamilton operator $H_{\rm eff}$ of the open system
provides, according to the boundary conditions, resonance or discrete states. 
The method for numerical calculations is given
in \cite{baroho,revopr} for nuclei, in
\cite{marost1} for atoms, in \cite{saro,datta} for quantum dots.

The individual states of the many-level system 
depend  on  parameters in a different  manner
according to their different spectroscopic properties. They
may therefore cross or avoid crossing as
discussed in sections \ref{eiv} and \ref{eif}. The most interesting effects
appear in the very neighborhood of the crossing points where
the contributions of all
the other states to the crossing phenomenon need not to be considered. 
Hence, the exceptional points defined in (\ref{int2}) and 
(\ref{int3}) play an important role  also in the many-level system.
For example, the
avoided crossing phenomenon of discrete and narrow resonance states 
is well known, see section \ref{avd}.

Most interesting is the population transfer related to an exceptional
point. The population transfer takes place not only at the crossing 
point itself but also in its neighborhood, i.e. at the critical point 
of an avoided level crossing. In realistic systems, population
transfer may be induced by means of lasers. In \cite{carla}, the
connection between high-order harmonic generation and the periodic
level crossings is investigated in detail. The knowledge of the
physical mechanism allows one to manipulate the adiabatic states and
consequently the harmonic spectra.  The results can  be
extended to a broader parameter range, as, for instance, those
characteristic of solid-state systems in strong fields.   
Another example studied recently, is the laser control of vibrational
transfer occurring at and in the neighborhood of exceptional points
\cite{atabek} where the resonances exchange their labels. It is
possible therefore to control, by means of a laser, the vibrational
transfer of the undissociated molecules from one field-free state to
another.

\section{The solution of the Schr\"odinger equation in 
the total function space } 
\label{sol}

The question arises now whether or not the properties discussed in the
foregoing sections survive when solving the  many-level problem
in the total function space.
In order to find an answer to this question let us sketch the 
Feshbach projection operator formalism \cite{feshbach}
that allows a {\it unified
description of structure and reaction}, for details see \cite{top}. 
The structure is determined by
the spectroscopic properties of the system ($Q$ subspace)  
while the reaction is induced by the environment of scattering
wavefunctions ($P$ subspace). 

The Schr\"odinger equation in the total function space reads 
\begin{eqnarray}
(H^{\rm full} -E)~\Psi^E_c & = & 0 
\label{sol1}
\end{eqnarray}
where 
\begin{eqnarray}
H^{\rm full} 
&\equiv & H_{QQ} + H_{QP} + H_{PQ} + H_{PP} 
\end{eqnarray}
is Hermitian and $H_{QQ} \equiv QHQ, ~H_{QP} \equiv QHP$ and so on. 
The two projection operators $Q$ and $P$ are defined by
\begin{eqnarray}
(H_{B}\!& -&\! E_k^{B}) \, \Phi_k^B = 0   
~\longrightarrow
~~Q  =  \sum_k|\Phi_k^B\rangle\langle \Phi_\lambda^B | 
\\
(H_{c}\!& -&\! E) \, \xi^{E}_{c} = 0  
~~~\longrightarrow
~~P  =  \sum_c\int_{\epsilon_c}^{\epsilon_{c}'} dE \; 
|\xi^{E}_{c} \rangle \langle \xi^{E}_{c} | 
\end{eqnarray}
and $P+Q=1$ is assumed, see section \ref{man}.
The operator $H_B=H^0_B+V$ contains the interaction $V$ between
the different basic states while the
coupling between the discrete and scattering  states is given by 
(\ref{man6}).
Thus, $\Psi^E_c$ contains (by definition) everything and 
$H^{\rm full}$ is Hermitian. 
The solution of the Schr\"odinger  equation 
in the total function space reads \cite{top}
\begin{eqnarray}
\Psi^E_c  =    \xi^E_c + 
\sum_{k,l=1}^N (\Phi_k^{B} +   \omega_k^0) 
\langle  \Phi_k^{B} | \frac{1}
{E-{H}_{\rm eff}} | \Phi_l^{B} \rangle
\langle   \Phi_l^{B} | H_{QP} | \xi^E_c \rangle 
\label{sol2}
\end{eqnarray}
where ${H}_{\rm eff}$ is given by (\ref{man3}) and
\begin{eqnarray}
 \omega_k^{0}  = G_P^{(+)}  H_{PQ} 
 \cdot  \Phi_k^{B} 
\; ; \quad G_P^{(+)}  = P (E - H_{PP})^{-1} P \; .
\label{sol20}
\end{eqnarray}
After diagonalizing ${H}_{\rm eff}$ (see equations (\ref{man7}) and
(\ref{man8})), the solution (\ref{sol2}) reads 
\begin{eqnarray}
\Psi^E_c =   \xi^E_c  ~+ \sum_{k=1}^N
\Omega_k \cdot  \frac{\langle \Phi_{k}^*| H_{QP} | 
\xi^E_{c} \rangle }{E - z_k}  \; .
\label{sol3}
\end{eqnarray}
Here 
\begin{eqnarray}
\Omega_k =(1+ G_{P}^{(+)} \; H_{PQ}) \, \Phi_k
\equiv (1+\omega_k)\, \Phi_k 
\label{sol4}
\end{eqnarray}
is the {\it wavefunction of the resonance state $k$}.
The tail of the resonance wavefunction is determined by $\omega_k$,
i.e. by a value analogue to (\ref{sol20}). 
The solution (\ref{sol3}) is exact in relation to the assumption $P+Q=1$. 

Of special interest is the  scattering wavefunction  
inside the localized part of the system.
According to (\ref{sol3}), it can be represented in a set
$\{\Phi_k \} $ of  biorthogonal wavefunctions,
\begin{eqnarray}
|\Psi^{E,R}_{c ~{\rm int}} \rangle & =& 
\sum_k c_{k E} |\Phi_k \rangle \; ; \quad
\langle  \Psi^{E,L}_{c ~{\rm int}} | = 
\sum_k c_{k E} \langle \Phi_k^* |
\label{sol13}
\end{eqnarray}
with the coefficients
\begin{eqnarray}
c_{k E} =  
\frac{\langle  \Phi_{k}^*| H_{QP} | \xi^E_{c} \rangle }{E - z_k}
\equiv \frac{1}{\sqrt{2\pi}} ~\frac{\gamma_k^c}{E-z_k} 
\label{sol14}
\end{eqnarray}
which  depend on energy. The coefficients $\gamma_k^c$ are
defined in (\ref{sol8}). Due to this representation, the 
phases of the wavefunctions $\Psi^{E}_{c ~{\rm int}}$ are not rigid. 
In analogy to (\ref{eif11}), the phase rigidity $\rho$ of the  
$\Psi^{E}_{c ~{\rm int}}$ with $1\ge \rho \ge 0$ can be defined \cite{top}.
It is possible therefore that some  wavefunctions 
align with the channel wavefunctions $\xi^E_c ~~(c=1,...,C)$.
This {\it alignment occurs by trapping other resonance states} \cite{top}:
finally, all but the aligned resonance states are more or less decoupled 
(trapped) from the continuum of scattering wavefunctions \cite{top}.

The resonance trapping phenomenon is proven  experimentally
\cite{stm}.
The alignment of resonance states to the scattering states of the environment 
is a collective phenomenon
to which all resonance states in a large energy region contribute,
see section \ref{dyn}.

\section{The $S$ matrix}
\label{sma}

The  $S$ matrix is given in the following manner \cite{ro91,top}
\begin{eqnarray}
S_{cc'}& = & 
\delta_{cc'} - \int \frac{\langle \chi^E_{c '} |V| \Psi^E_c 
\rangle }{E-E'} dE'
\nonumber \\
&=&
\delta_{cc'} - {\cal P} \int \frac{\langle \chi^E_{c '} |V| \Psi^E_c 
\rangle }{E-E'} dE' - 2 i \pi 
\langle \chi^E_{c'} | V | \Psi^E_{c} \rangle 
\nonumber \\ &\equiv & 
 \delta_{cc'} -  S_{cc'}^{(1)} -  S_{cc'}^{(2)} \; .
\label{sol5}
\end{eqnarray}
where the $\chi_c^E$ are the unperturbed scattering wavefunctions
and $\Psi^E_c$ is given in  (\ref{sol3}).
The $S$ matrix consists of two parts, one of which
\begin{eqnarray}
S_{cc'}^{(1)} & = &  
{\cal P} \int \frac{\langle \chi^E_{c '} |V| \Psi^E_c 
\rangle }{E-E'} dE' + 2 i \pi 
\langle \chi^E_{c'} | V | \xi^E_{c} \rangle 
\label{sol6}
\end{eqnarray}
depends smoothly on energy, and the other one
\begin{eqnarray}
 S_{cc'}^{(2)} &=& i \; \sqrt{2 \pi}  \; \sum_{k =1}^N
\langle \chi^E_{c'} | V | 
  {\Omega}_k \rangle    \cdot
  \frac{\gamma_{k}^{c}}{E - z_k}
\label{sol7}
\end{eqnarray}
is the  resonance term with the eigenvalues $z_k$  defined in
(\ref{man7}), and the coupling coefficients $\gamma_k^c$  defined in 
(\ref{sol8}).
If the resonance states are excited via the  continuum of decay channels 
it holds  
\begin{eqnarray}
\xi^E_c  = (1+G_P^{(+)}\cdot V)\; \chi^E_{c} 
\label{sol9}
\end{eqnarray}
and therefore
\begin{eqnarray}
\langle \chi^E_{c'} | V |{\Omega}_k \rangle
= \langle \xi^E_{c'} | V |{\Phi}_k \rangle 
\equiv \frac{1}{\sqrt{2\pi}}  ~\gamma_{k}^{c'}\; .
\label{sol10}
\end{eqnarray}
Using this relation, the resonance part of the $S$ matrix passes into the 
familiar expression
\begin{eqnarray}
S_{cc'}^{(2)} =  i \; \sum_{k =1}^N
  \frac{\gamma_{k}^{c}\;\gamma_k^{c'}}{E - z_k } \; .
\label{sol11}
\end{eqnarray}
However, there are some differences to the conventional expression of the 
$S$ matrix:
the coupling vectors $\gamma_{k}^{c}$ are calculated by means of the
eigenfunctions $\Phi_k$  of $H_{\rm eff}$ according to (\ref{sol8}),
the $z_k$ are  eigenvalues of  $H_{\rm eff}$, see (\ref{man7}), and
the  $\gamma_{k}^{c}$ as well as the  $z_k$ are energy dependent 
functions since $H_{\rm eff}$ depends explicitly on energy according to 
(\ref{man3}). Further,
the $S$ matrix is always unitary.

In the standard theory, the spectroscopic information is obtained 
from the poles of the $S$ matrix. According to (\ref{sol11}), this 
procedure is equivalent to $E=z_k$, i.e. the spectroscopic properties
are obtained from the eigenvalues $z_k(E)$ when $E$ is continued into
the complex plane. Hence, the double poles of the $S$ matrix contain the
information on the exceptional points. In the case
with two resonance states coupled to one channel $c$, 
the $S$ matrix at the crossing point  reads \cite{ro01}
\begin{eqnarray}
S&=& 1-i\sum_{k=1}^2 \frac{\gamma_k^c\gamma_k^c}{E-z_k} \nonumber\\
&=&1-2i\frac{\Gamma_d}{E-E_d+\frac{i}{2} \Gamma_d} -
\frac{\Gamma_d^2}{(E- E_d+\frac{i}{2} \Gamma_d)^2}
\label{sol12}
\end{eqnarray}
where  $ E_1=E_2\equiv E_d$, $ ~\Gamma_1=\Gamma_2\equiv \Gamma_d$
and (\ref{man7}) is used.
This expression shows a  non-linear behavior around the crossing
point. At the crossing point, the cross section vanishes due to
interferences \cite{mudiisro}. 
The interference minimum is however washed out 
in the neighborhood of the double pole \cite{ro03}. In any case,  
the resonance observed is broader than a Breit-Wigner resonance 
according to (\ref{sol12}). 
Different numerical studies are performed for realistic systems: for 
atoms in \cite{marost2,marost3,marost4} and for nuclei in \cite{plo}. 
In all cases, the dependence of, e.g., the cross section on a 
certain parameter is nonlinear in the neighborhood of crossing points 
\cite{plo,marost2,marost3}.

As a result, it can be stated that the effects induced by the
exceptional  points  survive when the problem
in the total function space is considered and observables are calculated.

\section{The entanglement of  states}
\label{mix}

The eigenfunctions $\Phi_i$ of the non-Hermitian Hamiltonian $\heff$ 
can be represented in relation to different sets
of basic wavefunctions: 
\begin{enumerate} 
\item
Representation of the $\Phi_i$ in the $\{\Phi_n^0\}$, 
\begin{equation}
\Phi_i=\sum_{j=1}^N \, b_{ij} ~\Phi_j^0 \; ;
 \quad \quad b_{ij} = \langle \Phi_j^{0 *} | \Phi_i\rangle  \; ,
\label{mix4}
\end{equation}
where the $\Phi_i^0$ are  eigenfunctions 
of the non-Hermitian unperturbed operator $H^0_{\rm eff}$
(with vanishing non-diagonal matrix elements),
\begin{eqnarray}
(H^0_{\rm eff}  - z_i^0) ~| \Phi_i^0 \rangle  =  0 \;.
\label{mix3}
\end{eqnarray}
\item
Representation of the $\Phi_i$  in the $\{ \Phi_n^B \}$ 
\begin{equation}
\Phi_i=\sum_{j=1}^N \, a_{ij} ~\Phi_j^B \; ;
 \quad \quad a_{ij} = \langle \Phi_j^B | \Phi_i\rangle  
\label{mix6}
\end{equation}
where the $\Phi_i^B$ are eigenfunctions 
of the Hermitian operator $H_B = H^0_B + V$, 
\begin{eqnarray}
( H_B  - E_i^B) ~| \Phi_i^B \rangle  =  0 \; .
\label{mix5}
\end{eqnarray}

\end{enumerate}
When the $\Phi_i$ and the
wavefunctions of the basic set describe localized non-overlapping states, the 
representations (\ref{mix4}) and (\ref{mix6}) are linear and well defined. In
such a case,
the number of states described by the $\Phi_i$ is equal to the number of 
states described by the $\Phi_i^0$ and $\Phi_i^B$, respectively. 

The situation is more complicated when the mixing of the wavefunctions
$\Phi_i$  in the overlapping regime is considered. The reason is the fact that
the number of states is reduced at the exceptional points. 
Here and at the critical points of avoided level crossings, respectively,  
width bifurcation starts. As a result, some of the states become delocalized
(short-lived) and do no longer contribute to the number of 
localized  (long-lived) states. Hence, the number $N^{\rm loc}$ of
narrow (localized) resonance states described by the 
$\Phi_i$ may be different from the number $N$ of basic states described 
by the $\Phi_i^0$ and $\Phi_i^B$, respectively. In spite of 
$N^{\rm loc} \ne N$, a representation of the wavefunctions
$\Phi_i$ in the set $\{\Phi_n^0\}$ of $N$ wavefunctions is formally possible.
However, the spectroscopic linear relation between the long-lived
localized eigenstates   and the basic   localized  
resonance states   (or the basic localized discrete 
states)  is lost. For the coefficients $b_{ij}$ holds
\begin{eqnarray}
\delta_{i,j} = ~\langle \Phi_i^* | \Phi_j \rangle 
= \sum_{k,l=1}^N b_{ik}b_{jl} ~\langle \Phi_k^{0*}|\Phi_l^{0}\rangle 
= \sum_{k=1}^N b_{ik}b_{jk} \; . 
\label{mix8}
\end{eqnarray}

Numerical studies for the nuclear reaction $^{15}F+p$ have shown the
strong energy dependence of the coefficients $b_{ij}$  around the critical
point of avoided and true crossings of eigenvalue trajectories \cite{plo}.  
They show the
exchange of the two  resonance states at the critical point of
the avoided level crossing (characterized by level repulsion); 
they become infinitely large at the crossing point, and depend 
resonance-like on energy  at the crossing point of the energy
trajectories after width bifurcation. This picture agrees with that
obtained for laser-induced continuum structures in atoms
\cite{marost2,marost3}, and also with that received from a 
study of the $2\times 2$ toy model \cite{ro01}.

According to these results, the
scenario in the regime of overlapping resonances is the following. 
At a (true or avoided)  crossing  point of the trajectories 
of two resonance states,
one of the resonance states starts to align to a scattering state of the 
environment and loses its localization while
the other one remains localized  but loses also its 
spectroscopic relation  to the basic localized states.
In a many-level system, this scenario repeats hierarchically 
\cite{top,isrodi}. Finally,
the number of localized long-lived states is reduced and,
moreover, the surviving narrow resonance states -- 
although (almost) localized --
have lost their spectroscopic relation to the basic individual 
localized states. They are strongly entangled.
This scenario is called  {\it dynamical phase transition}.

Dynamical phase transitions are observed in different experimental
studies. Some of them will be sketched in section \ref{exp}. Here, it
will be underlined only that -- according to the above discussion --  
dynamical phase transitions can be traced
back to the existence of exceptional points in quantum
systems and to the nonlinearities caused by them.
Moreover, the $\omega$ in (\ref{int3}) and 
(\ref{mix1}) contain solely the
coupling of the states via the environment. The dynamical phase
transition is  therefore
an {\it environmentally induced phenomenon}.

\section{The avoided crossing phenomenon of discrete and narrow 
resonance states}
\label{avd}

According to section \ref{man}, two states may interact via the 
environment of scattering
wavefunctions even if their energy  is outside the energy window
coupled directly to the environment of scattering wavefunctions.
Numerical calculations have shown that the avoided  crossings of
discrete states can be traced back, indeed, to the crossing points of 
resonance states by varying one or two parameters \cite{ro01}. The
model Hamiltonian used in the calculations is 
\begin{eqnarray}
H=\left(
\begin{array}{cc}
e_1^0(a) & 0 \\
0 & e_2^0(a) 
\end{array} \right)  
- \left(
\begin{array}{cc}
i/2~\gamma_1^0 & \omega \\
\omega & i/2~\gamma_2^0 
\end{array} \right)   
\label{avd4}
 \end{eqnarray}
with the notation $\epsilon_k \equiv e_k^0 - \frac{i}{2} \gamma_k^0$.
The calculations are performed as a function of the parameter $a$ 
for different $\gamma_1^0$ (and fixed ratio $\gamma_1^0/\gamma_2^0$ and
fixed $\omega$) such that -- with decreasing $\gamma_1^0$ -- 
the levels (i) have always different widths and
cross freely in energy at the  value $a=a^{\rm cr}$, 
(ii) cross in energy and width at $a=a^{\rm cr}$ when
$\gamma_1^0=\gamma_1^{\rm cr}$, 
(iii) avoid crossing in energy when the widths cross at $a=a^{\rm cr}$
and (iv) pass into discrete states (with vanishing width)
and avoid crossing at $a=a^{\rm cr}$ (in a similar manner as the 
narrow resonance states with $\gamma_1^{0} < \gamma_1^{\rm cr}$ do).

Most interesting results are obtained for the coefficients
$|b_{ij}|^2$ defined in (\ref{mix4}). At the critical value $a^{\rm  cr}$, 
it is $\delta \equiv |b_{i,j=i}|^2 -|b_{i,j\ne i}|^2 =1$  when the two levels
cross freely in energy (for $\gamma_1^0 > \gamma_1^{\rm cr}$) 
according to the fact that the two states 
exist at different time scales and are therefore well defined.
Approaching the crossing point (at $\gamma_1^0 = \gamma_1^{\rm cr}$),  
~$|b_{i,j=i}|^2$ and 
$|b_{i,j\ne i}|^2$ increase up to $\infty$ and $\delta \to 0$ what 
is achieved due to (\ref{eif4}). When the two states avoid crossing
(for $\gamma_1^0 < \gamma_1^{\rm cr}$), 
~$|b_{i,j=i}|^2$ and $|b_{i,j\ne i}|^2$ decrease, however 
$\delta = 0$ remains at the critical point $a^{\rm cr}$. This result is an
expression for the fact that the two
states are exchanged at the critical point. Also for discrete
states ($\gamma_1^0 =0$), 
~$\delta =0$ at the critical point $a^{\rm cr}$ of avoided level crossing. 
Here, $|b_{i,j=i}|^2 =|b_{i,j\ne i}|^2 = 1/2$. 

It is interesting to see that $|b_{i,j=i}| \ne 1$ and $|b_{i,j\ne i}| \ne 0 $
in a comparably large parameter
range around the critical point. This range 
is the larger the smaller the widths of the states are
(when  traced as a function of the parameter $a$ as in the above example). 
The range  shrinks to one point (the crossing point) when the 
two states cross, i.e. when the two states overlap completely.
The largest range occurs for discrete states although the discrete states do
not overlap. This means that also discrete states 
are mixed via the continuum in a finite range  $\Delta a$ of the 
parameter $a$ and that this mixing is caused by the existence of the
exceptional point. 

In any case, resonance states as well as discrete states are 
mixed in a finite range $\Delta a$ of the parameter $a$ 
around the critical point $a=a^{\rm cr}$ of an avoided crossing.
At high level density  where many neighboring levels avoid crossing,
the ranges $\Delta_i a$ corresponding to a
non-vanishing mixing of two states in each case, may overlap. 
As a consequence, the eigenfunctions $\Phi_i$ of $\heff$ that
describe localized states,  lose their spectroscopic
relation to the wavefunctions $\Phi_i^0$ of the basic individual states
in a certain range of the parameter $ a$  which is determined by the 
sum of the overlapping $\Delta_i a$.
This statement holds true also for discrete states. It agrees fully
with the conclusion drawn in section \ref{mix} on the dynamical phase
transition that occurs in the regime with many avoided level crossings.

Thus, the strong mixing of the wavefunctions at high level density  
expressed by the lost of their spectroscopic relation to the individual 
basic resonance states is, on the one hand, a hint to the existence of
exceptional points in the continuum of scattering wavefunctions. 
On the other hand, it points to the dynamical phase transition
related to the many avoided crossings of localized states and the
nonlinear effects caused by them.

The strongly mixed trapped (localized) states can be described best 
by statistical methods, e.g. by the Gaussian orthogonal ensemble.
An example is  the nuclear data ensemble analyzed by 
Bohigas et al. \cite{bohigas} many years ago. It caused many studies
on quantum chaos. The states of the Gaussian orthogonal ensemble 
are different from those of a two-body random ensemble. They
do not decay according to an exponential law \cite{hardit}. This points 
to the fact that they differ from individual resonance states 
in spite of their small widths preventing them from
overlapping (level repulsion is one of the characteristics of the Gaussian
orthogonal ensemble).  Indeed, the long-lived chaotic states in nuclei coexist
with a broad single-particle resonance by which they are overlapped.
For details see \cite{jmp}.
In  \cite{shot} a shot noise analysis of the states of a microwave billiard
is performed.  
In the long-time scale, the system shows features characteristic of quantum
chaos while the system is regular in the short-time scale. In a recent study,
the statistical properties of lifetimes of electromagnetic quasibound states
in dielectric microresonators with fully chaotic ray dynamics are discussed
\cite{schom}. The results show regular short-lived resonances and many
long-lived resonances. The level statistics of the last ones 
is very well described by a random-matrix model, provided that two effective
parameters are appropriately renormalized. This renormalization is linked 
by the authors  \cite{schom} to
the formation of the short-lived resonances, i.e. to the resonance-trapping
phenomenon. 

All these studies show, on the one hand, a correlation between 
exceptional points and quantum chaos. On the other hand, 
they show that quantum chaotic states
are different from individual resonance states which characterize the system
at low level density. In other words, quantum chaotic states are the
result of a dynamical phase transition induced by the environment 
(including boundary conditions).   They coexist with a few 
(aligned) short-lived states.

\section{The interplay between system and environment}
\label{dyn}

Some years ago, the question has been studied \cite{jung}  whether or not
the resonance trapping phenomenon is related to some type of phase 
transition. 
The study is performed by using the toy model 
\begin{eqnarray}
H_{\rm eff}^{\rm toy} = H_0 + i \alpha VV^+
\label{sol15}
\end{eqnarray}
in the one-channel case and with the assumption that (almost) all 
exceptional points  accumulate in one point \cite{hemuro}.
It has been found that  resonance trapping  
may be understood, in this case, as a second-order phase transition.
The calculations are performed for a linear chain consisting of a
finite number $m=2n+1$ of states. The state in the center of the spectrum 
traps the other ones and becomes a collective state in a global
sense: it contains components of almost all basic states of the
system, also of those which are not overlapped by it. The normalized width 
$\Gamma_0/m$ of this state can be considered as the order parameter:  
it increases linearly as a function of $\alpha$, and the first
derivative of $\Gamma_0/m$ jumps at the critical value 
$\alpha = \alpha^{\rm cr}$.    
The two phases differ by the number of localized states. In the case
considered, this number is $m$ at $\alpha < \alpha^{\rm cr}$, 
and $m-1$ at  $\alpha > \alpha^{\rm cr}$. 

Much more interesting is the realistic case with the Hamiltonian 
(\ref{man3}). Here,  trapping of resonance states occurs 
in the regime of overlapping resonances hierarchically, i.e.
one by one \cite{isrodi}. The crossing points do 
{\it not} accumulate in one point, but are distributed over a certain 
range of the parameter. In this case, a {\it dynamical} 
phase transition takes place in a finite parameter range 
inside the regime of overlapping resonances \cite{top}. 
Also in this case, almost all resonance states are involved in the phase
transition and, furthermore, the number $N$ of localized states is
reduced. That means, the $Q$ subspace splits into two parts under the
influence of the environment. One part contains the few short-lived
states which are (more or less) aligned
to the scattering states of the environment, while
the other part contains the trapped, long-lived and well localized states. 
Both time scales are well separated from one another. 

An example are the short-lived whispering gallery modes in a microwave cavity
with convex boundary which coexist  with many long-lived states 
\cite{shot,naz1,local}. Another example is known in 
nuclear physics: the short-lived single-particle 
resonances are  responsible for the fast direct reaction part, 
while the long-lived ones cause the slow compound-nucleus 
reaction part. In the Feshbach {\it unified theory of nuclear reactions} 
\cite{feshbach}, the direct reaction part is described exactly while the
compound-nucleus reaction part is described by means of statistical ensembles.
Similar representations are used in other fields of physics at high level
density. 

Interesting is the enhancement of observable values 
in the parameter range in which the phase transition takes place. 
The enhancement is a direct consequence
of the  alignment that occurs in such a 
manner that the aligned state fits best to the environment, i.e. that
the corresponding $\gamma_k^c$ is maximal. An example is 
studied theoretically in \cite{burosa}. Here, an
anticorrelation between the conductance $|t|^2$ of a quantum dot and 
the phase rigidity $|\rho|^2$ is found. 
The alignment is basic for the solution of the brachistochrone problem in
quantum mechanics \cite{top,brach}.

In any case, the regime 
at low level density (or small coupling via the continuum)  
differs from the regime at high level density (corresponding to 
strong coupling via the
continuum). At small coupling via the continuum,   the resonance states 
show individual spectroscopic features which  are lost at large coupling. 
Here, many narrow ({\it trapped}) resonances
are superimposed on broad ({\it aligned}) resonances. The trapped resonance
states show chaotic features (section \ref{avd}).

The dynamical phase transitions are surely the most interesting feature of 
non-Hermitian quantum physics. They are environmentally induced, 
see section \ref{mix} and \cite{jopt}. 
Mathematically, this phenomenon is directly related to the existence of 
exceptional  points, i.e. to the coupling matrix elements
$\omega$ in (\ref{int3}),  to the phase rigidity $r_k$ of the eigenfunctions 
and to the nonlinear terms in the Schr\"odinger equation, see
(\ref{mix1}) and (\ref{mix2}).  In detail:
\begin{itemize}
\item[(i)]
The phases of the eigenfunctions of the non-Hermitian Hamilton operator are
not rigid in approaching the exceptional  point: $r_k < 1$ in the
regime of overlapping resonances. 
\item[(ii)]
Due to $r_k<1$, some resonance states may align with the scattering
states of the environment while other ones decouple from the environment
(width bifurcation). 
\item[(iii)]
The short-lived aligned resonance states lose, to a great deal, their 
localization and make the system (almost) transparent. 
\item[(iv)]
The long-lived trapped resonance states are well localized and
show chaotic features.
\item[(v)]
The spectroscopic relation between the localized states at low level density 
(without resonance overlapping)  and  those at high level density 
(with  overlapping short-lived and long-lived resonances) is lost.
\end{itemize}

The appearance of dynamical phase transitions 
can explain some puzzles that are observed experimentally and could not 
explained theoretically in the framework of conventional 
Hermitian quantum theory. Some experimental results of such a type 
will be sketched in the following section \ref{exp},
together with experimental results which prove directly the 
dynamical phase transition.

\section{Dynamical phase transitions in experimental data}
\label{exp}

\subsection{Experimental verification of the resonance trapping 
phenomenon}

About 10 years ago, the first direct experimental verification of
the counterintuitive  resonance trapping phenomenon is 
presented \cite{stm}. The experiment is based on the equivalence of
the electromagnetic spectrum for flat cavities to the quantum
mechanical spectrum of the corresponding system. This equivalence
holds also when the system is  opened by
coupling the discrete states of the cavity to an attached waveguide.
In the experiment \cite{stm},  a microwave Sinai cavity with an
attached waveguide with variable slit width was used. 

As a result of this experimental study, agreement with theory is observed:
the widths of all resonance states first
increase with increasing coupling strength to the channels (continuum
of scattering wavefunctions)  but
finally decrease again for most of the states. Thus,
the dynamical phase transition has been directly traced in 
this experiment.

\subsection{Spectra of light and heavy nuclei}
\label{nucl}

It is a well-known result of nuclear physics studies during many years
that the resonance states in light nuclei are different from those in heavy
nuclei. In light nuclei,  resonance states appear
mostly at low excitation energy of the nucleus, where the level density
is small. The lifetimes of the resonance states are 
often near to the limit for single-particle (or alpha) decay.  
All resonance states show individual spectroscopic features. 

The situation in heavy nuclei is completely different. The first (elastic) 
threshold for particle decay is at about 8 MeV 
excitation energy of the nucleus where the level
density is extremely high. In a small energy region above this threshold, the
so-called neutron (compound nucleus) resonances are identified. 
They are extremely long-lived corresponding to decay widths of the 
order of eV and show chaotic features \cite{bohigas}, see section 
\ref{avd}. Much less discussed in literature are
the so-called single-particle resonances in heavy nuclei the widths of which
are of the order of magnitude of MeV. Their energy is mostly just below the 
elastic decay threshold and their width at energies above the threshold 
(see section \ref{sma} for the energy dependence of the widths)
is much larger than the widths of the long-lived states. 
In the cross section, they appear as a smooth
background for the very narrow neutron resonances. The time scales of these 
two different types of resonance states are well separated from one another: 
up to $10^6$ neutron resonances are overlapped by  one single-particle 
resonance.  

In medium-mass nuclei, the first (elastic) decay threshold
is at a comparably low excitation energy of the nucleus where  
the level density is still relatively low. These nuclei
are characterized by  overlapping resonances with different lifetimes. 
They are described well by the {\it  doorway picture}
according to which doorway states 
coexist with long-lived compound nucleus resonance states. The doorway 
states being comparably short-lived, are coupled 
directly to the decay channels {\it  and} to the narrow 
compound nucleus resonance states.
The  narrow resonance states, however, are assumed to be coupled to the
continuum {\it only} via the doorway states.  This model 
gives a good description of medium-mass nuclei, see \cite{jmp}. In 
\cite{auerbach}, the doorway picture is related to the Dicke model 
super-radiant mechanism
which is nothing but the mechanism of resonance trapping (see \cite{soze}
and section \ref{sol}). 

In \cite{plo}, exceptional points are identified in nuclei under realistic 
conditions. This allows us to consider nuclei at low and high level density 
as quantum systems, respectively, below and above a dynamical phase 
transition. The phase transition itself can not be controlled today 
by means of a parameter
since the strong nuclear forces do not allow a manipulation of nuclei. 
In \cite{jmp}, the resonance states at high level density
are shown to be trapped states, i.e. states originating from a
dynamical phase transition. They are described well by a statistical 
ensemble containing the interaction between {\it all} particles
(e.g. by the Gaussian orthogonal ensemble), and not by a two-body ensemble.  
Beyond a critical value, the widths of these
states decrease with increasing coupling strength between system and
environment (continuum of scattering wave functions) \cite{diharo}. 
The states of the Gaussian orthogonal ensemble decay according to a 
power law \cite{hardit}.

\subsection{Phase lapses}
\label{laps}

In  experiments \cite{heiblum1,heiblum2} 
on Aharonov-Bohm rings containing a quantum dot in one arm, 
both the phase and the magnitude of the transmission amplitude
$T=|T|~e^{i\beta}$ of the dot can be extracted. The obtained results 
caused much discussion since they do not fit into the standard 
understanding of the transmission process.  
As a  function of the plunger gate voltage $V_g$, a series of well-separated 
transmission peaks of rather similar width and height has been observed
in many-electron dots
and, according to expectations, the transmission phases $\beta(V_g)$ 
increase continuously by $\pi$ across every resonance. 
In contrast to expectations, however, 
$\beta$ always jumps sharply downwards by $\pi$  in each valley
between any two successive peaks. These jumps called phase lapses, 
were observed 
in a large succession of valleys for every  many-electron dot studied.
Only in few-electron dots, the expected so-called mesoscopic behavior
is observed, i.e. the phases are sensitive to details of the dot 
configuration. The problem is considered theoretically, 
in the framework of conventional Hermitian quantum physics, in many papers 
over many years, however without solving it.

In  \cite{murophas},  the phase lapses observed experimentally at high
level density are related to the trapped resonance states resulting from the
dynamical phase transition. In accordance to this picture, only
the resonance states at low level density  show  
individual spectroscopic features. At high level density, the observed
resonances arise from trapped states. They 
show level repulsion (see section \ref{avd}) and have almost
no spectroscopic relation to the open decay channels such that phase
lapses appear. It follows further,
that any theoretical study on the basis of conventional Hermitian
quantum physics is unable to explain the experimental results
convincingly. In other words: the experimentally observed phase lapses can be
considered to be a proof for the dynamical phase transitions occurring in 
mesoscopic systems.

\subsection{Dephasing at very low temperature}
\label{deph}

Comparing the basic ingredients of the theory 
of open quantum systems with the
experimental results on dephasing at very low temperature, 
it should firstly be stated that the concept dephasing is used 
differently in  different papers. 
Here, we consider the phase coherence time $\tau_\phi$ characterizing 
dephasing at very low temperature. In the following, a very
short discussion of the results obtained experimentally will be given.
The discussion is
qualitatively by using the  results obtained in different recent studies.
It avoids to comment the many controversial discussions  that exist
in the literature to this question. 

In the proceedings  of a recent conference,
the experimental progress on the saturation problem in metallic
quantum wires is reviewed \cite{proceedings1}. 
As a conclusion of this analysis, based on all
presently available measurements of the phase coherence time $\tau_\phi$
in very clean
metallic wires, it is hard to conceive that the apparent saturation of
$\tau_\phi$ is solely due to the presence of an extremely small amount of
magnetic impurities.  

The absolute value of $\tau_\phi$ (and not just its temperature
dependence) is studied in  \cite{hackens}. It is found that
the electron dwell time $\tau_d$ is the central parameter
governing the saturation of phase coherence at low temperature. The condition
for the occurrence of saturation is found to be $\tau_{\phi}^{\rm sat} \approx
\tau_d$ where  $\tau_{\phi}^{\rm sat}$ is the saturated coherence time.
This simple behavior holds over the three orders
of magnitude covered by the available data in the
literature. According to the authors, $\tau_\phi$ is found to be 
intrinsic to the physics of the quantum dots, and
not due to the coherence time of the electrons themselves.
Furthermore it is found \cite{hackens} 
that $\tau_\phi$ is strongly influenced by the
population of the second electronic subband in the quantum well. 

According  \cite{proceedings2},
one consensus has been reached by several groups, saying that the 
responsible electron dephasing processes in highly disordered
and weakly disordered metals might be dissimilar. That means, while 
one mechanism is responsible for dephasing
in weakly disordered metals, another mechanism may be relevant
for the saturation (or very weak temperature dependence) of
$\tau_\phi$  found in highly disordered alloys. According to the
authors of \cite{proceedings2}, the intriguing electron
dephasing is very unlikely due to magnetic scattering. It may
originate from specific dynamical structure defects in the samples.   

Experimental data from  many different publications for $\tau_\phi^0$
obtained in metallic samples with different diffusion coefficients, 
are collected in \cite{proceedings3}.
The conclusion is that low temperature saturation of
$\tau_\phi$ is universally caused by electron-electron interactions.  
The authors found seemingly contradicting dependencies of $\tau_\phi^0$ on the
diffusion coefficient $D$ in weakly and strongly disordered conductors. While
the trend {\it less disorder -- less decoherence} for sufficiently clean
conductors is quite obvious, the opposite trend {\it more disorder -- less
decoherence} in strongly disordered structures is unexpected.  

All these statements obtained from the results of many experimental studies 
fit qualitatively
into the expectations received by considering the quantum dot as an open
quantum system. First of all, the saturation  of $\tau_\phi$ appears 
in a natural manner since most states of an open quantum dot have a finite
lifetime at zero temperature. The value of the lifetime can be obtained 
from the imaginary part of the 
complex eigenvalue $z_\lambda$ of the non-Hermitian Hamilton operator 
$H_{\rm eff}$ [i.e. from Im$(z_\lambda)$]. It expresses the time the electron 
stays in the quantum dot. This time is called usually dwell time. Thus,
the result obtained in \cite{hackens} supports 
the description of the quantum dot as an open quantum system.

Also the more complicated result of different processes in weakly and strongly
disordered systems is by no means in contradiction to the properties
known for the eigenstates of open quantum systems. 
In some cases, $\tau_\phi^0$ depends only weakly on the
electron diffusion constant $D$: it is somewhat smaller when $D$ is larger. 
That means, states with a large lifetime give only a small 
contribution to the diffusion -- a result
which is very well known.  In other cases,
the relation between $\tau_\phi^0$ and the diffusion constant $D$ shows the
opposite trend. Also in this case the states with a large lifetime give,
of course, a small contribution to the diffusion. In contrast to the foregoing
case, however, the   main contribution to the diffusion 
arises obviously from short-lived  states (according to
the resonance trapping phenomenon). 
Finally, the short-lived states form some background for the 
long-lived resonance states.  The diffusion constant is
determined mainly by the contribution of the background states. Therefore, the
diffusion constant $D$ increases with increasing   $\tau_\phi^0$ of the
(long-lived) resonance states -- a  result being counterintuitive in the same
manner as the resonance trapping effect.  The last one is directly proven
experimentally \cite{stm}.

In this respect another experimental result obtained in
\cite{hackens} is interesting. It shows that, in the systems considered,  
the quantity  $\tau_\phi$ is strongly influenced by the
population of the second electronic subband in the quantum well. 
Obviously this means that the degree of overlapping of the states
plays an important role for the lifetimes of the states
-- according to one of the basic properties 
of the eigenstates of $H_{\rm eff}$. 
Further experimental studies related to this question  would be very useful.

As a result of this discussion:
dephasing shows features that might be related to the
non-rigidity of the phases of the wavefunctions
of an open quantum system and to the dynamical phase transition occurring
in the regime of overlapping resonances. Accordingly, 
the coherence time  $\tau_\phi$ is intrinsic to the physics of 
the quantum dot, and not due to the coherence time of the electrons.
This conclusion agrees qualitatively with that obtained from the 
experimental results.
A quantitative description of the experimental data by using the 
theory of open quantum systems with a non-Hermitian Hamilton
operator, is not performed up to now. It is, however, interesting to
remark that a decoherence rate $1/\tau_\phi$ appears also in the
dynamics of a spin swapping operation where it is well defined
(section \ref{spin}).

\subsection{Quantum dynamical phase 
transition in the spin swapping operation}
\label{spin}

A swapping gate in a two-spin system exchanges the degenerate states 
$|\uparrow , \downarrow \rangle $ and $|\downarrow ,\uparrow \rangle $.  
Experimentally, this is achieved by turning on and off the spin-spin
interaction $b$ that splits the energy levels and induces an oscillation
with a natural frequency $\omega$. An interaction  $\hbar /\tau_{SE}$ 
with an environment of neighboring spins degrades this oscillation
within a decoherence time scale $\tau_\phi$. The experimental
frequency $\omega$ is expected to be roughly proportional to 
$b/\hbar$ and the decoherence time  $\tau_\phi$ proportional to
$\tau_{SE}$. In \cite{pastawski}, experimental data are presented
that show drastic deviations in both $\omega$ and $\tau_\phi$ from
this expectation. Beyond a critical interaction with the environment,
the swapping freezes and the decoherence rate drops as $1/\tau_\phi
\propto (b/\hbar )^2 \tau_{SE}$. That means, the relaxation
decreases when the coupling to the environment increases.
The transition between these two
quantum dynamical phases occurs when $\omega  \propto
\sqrt{(b/\hbar)^2 -(k/\tau_{SE})^2}$ becomes imaginary (where $k$
depends only on the anisotropy of the system-environment interaction,
$0 \le k \le 1$). The experimental results are interpreted 
by the authors as an environmentally induced quantum dynamical
phase transition occurring in the spin swapping operation 
\cite{pastawski,swap}.

Further theoretical studies within the Keldysh formalism 
showed that $\tau_{\phi}$ is a non-trivial function of the 
system-environment interaction rate $\tau_{SE}$, indeed: it is 
$1/\tau_\phi \propto 1/\tau_{SE}$ at low  $\tau_{SE}$ 
(according to the Fermi golden rule) but
$1/\tau_\phi \propto \tau_{SE}$ at large $\tau_{SE}$. This theoretical 
result is in (qualitative) agreement with the experimental results.
In \cite{swap2}, the dynamical phase transition 
in the spin swapping operation is related to the
existence of an exceptional point. 

The dynamical phase transition observed experimentally in the spin 
swapping operation and described theoretically within the Keldysh formalism
shows qualitatively the same features as the dynamical phase transitions
discussed in the present paper on the basis of the resonance trapping
phenomenon (width bifurcation).

\subsection{Loss induced optical transparency in complex optical 
potentials}
\label{opt}

Recently, the prospect of realizing complex $\cpt$ symmetric
potentials within the framework of optics has been suggested 
\cite{makris1,makris2,makris3}. It is based on the fact that the
optical wave equation is formally equivalent to the quantum 
mechanical Schr\"odinger equation.
One expects therefore that $\cpt$ symmetric optical lattices 
show a behavior which is qualitatively similar to that
discussed  for open quantum systems in the present paper. 

Experimental studies showed, indeed, a phase transition that leads to
a loss induced optical transparency in  specially designed
non-Hermitian guiding potentials \cite{guo,nature}: 
the output transmission first decreases, attains a minimum
and then increases with increasing loss. The phase
transition is related, in these papers, to  $\cpt$ symmetry breaking.
In a following theoretical paper \cite{makris4}, the Floquet-Bloch
modes are investigated in $\cpt$ symmetric complex periodic
potentials. As a result, the modes are skewed (nonorthogonal) and
nonreciprocal. That means, they show the same features as  modes 
of an open quantum system under the influence of exceptional points.
A detailed discussion of this analogy is given in \cite{jopt}.
The optical realization of relativistic non-Hermitian quantum
mechanics is considered in \cite{longhidirac}. Here, the $\cpt$
symmetry breaking of the Dirac Hamiltonian is shown to be related to
resonance narrowing what is nothing but resonance trapping.

The title of one of the papers published in Nature Physics
\cite{nature} to this topic reads: {\it Broken symmetry makes light work}. 
It is exactly this property which characterizes the  phase
transition in complex optical potentials. However,
the situation in open quantum
systems is qualitatively the same: in the dynamical phase transition, 
the spectroscopic relation to the individual resonance
states at low level density (including all  symmetries)
is broken and the system becomes transparent, see e.g. section \ref{dyn}.

\section{Summary}
\label{sum}

In the present paper, exceptional points are shown to be responsible
for mainly two  properties of quantum systems. Both  
condition one another.

First, the spectroscopy of discrete and resonance states is 
strongly influenced by exceptional points in their neighborhood.
Both types of states are  eigenstates of one and the same 
(non-Hermitian) Hamilton 
operator, but the boundary conditions differ from one another. 
The states are discrete (corresponding to an infinite long lifetime) 
when their energy is beyond the window coupled
to the continuum of scattering wavefunctions.
The states are resonant (corresponding, in general, to a finite lifetime) 
when their energy is inside the window coupled to the continuum of
scattering wavefunctions. Accordingly,
the singularities (crossing points) in the continuum influence not only the
behavior of resonance
states but also that of discrete states. 

Discrete states are described well in the framework of conventional 
quantum mechanics as known for very many  years, although it is
necessary to introduce effective forces in the conventional theory 
(which arise, at least partly, from the principal value 
integral  of the coupling term via the continuum).
The Hamiltonian is Hermitian and $A_k = 1 $,
the phases of the eigenfunctions are  rigid 
corresponding to $r_k \equiv A_k^{-1} = 1 $,
the discrete states avoid crossing and
the topological phase of the {\it diabolic point} 
is the Berry phase. Due to $A_k = 1 $, 
the Schr\"odinger equation is linear, but
the levels are mixed (entangled) in the total parameter range of 
avoided level crossing. At the critical point,
the mixing is maximal (1:1). 

Resonance states are described well when the  quantum theory
is extended by including the environment of scattering wavefunctions.
The Hamiltonian is, in general, non-Hermitian and $A_k \ge 1 $,
the phases of the eigenfunctions are, in general, not rigid 
corresponding to $0 \le r_k \equiv A_k^{-1} \le 1 $,
the resonance states can cross in the continuum and
the topological phase of the {\it crossing point} is twice the Berry
phase. When $0 < r_k \equiv A_k^{-1} < 1 $
(regime of resonance overlapping and avoided level crossings), 
the Schr\"odinger equation is nonlinear and
the levels are strongly mixed (entangled). 
The parameter range in which mixing appears, shrinks to one point 
when the levels cross, i.e. when $r_k \equiv A_k^{-1} \to 0$.

Secondly, dynamical phase transitions are caused by exceptional points.
According to the results given in the present paper, a dynamical 
phase transition occurs in the regime of overlapping resonances.
It is produced by width bifurcation, is environmentally induced
and breaks spectroscopic symmetries characteristic of the system.
It consists in the reduction of the number of localized states by 
alignment of a few resonance states to the (extended) scattering states. 
By this, it breaks the spectroscopic relation between  states below and 
beyond the dynamical phase transition.

The two phases below and beyond the dynamical phase transition
are characterized by the following properties.
In one of the phases, the discrete and narrow resonance states have
individual spectroscopic features. Here, the real parts (energies) of the
eigenvalue trajectories  avoid crossing while 
the imaginary ones (widths) can cross.
In the other phase, the narrow resonance states are superimposed 
with a smooth background and the individual spectroscopic features 
of the states are lost. The narrow resonance states and, respectively, 
the corresponding discrete states show chaotic features.
They do not cross in energy, but show level repulsion.
The real parts (energies) of the eigenvalue trajectories of 
narrow resonance states can cross with those of the broad states since 
the narrow and broad states exist at well separated time scales.
In the transition region, the different time scales
corresponding to the short-lived  and 
long-lived resonance states are formed, and the overlapping of the 
different resonance states is directly visible in the cross section. In this 
regime, the cross section is enhanced due to the (at least partial) alignment 
of some states with the scattering states of the environment.

It is interesting to see that the system behaves according to expectations
only at low level density. After passing the transition region with
overlapping resonances by further variation of the parameter, the behavior of
the system becomes counterintuitive: the narrow resonance states decouple more
or less from the continuum of scattering wavefunctions and the number of
localized states decreases. 

According to the results represented in the present paper,
the role of exceptional points in quantum physics can
be seen best in the non-Hermitian quantum physics. 
Knowing the mathematical properties of the exceptional points
it is possible, on the one hand, to explain (qualitatively)
some experimental results which could not be understood in the 
framework of the conventional 
Hermitian quantum physics in spite of much effort.
Numerical calculations for some realistic cases have to be performed  
in order to compare theory and experiment in detail. 
On the other hand, quantum systems can be manipulated 
systematically for applications.
Another interesting topic of non-Hermitian quantum physics results from
the formal equivalence of the optical wave equation in $\cpt$ symmetric 
optical lattices to the quantum mechanical Schr\"odinger equation.
This equivalence allows to receive much new information on quantum systems. 

In any case, further theoretical and experimental studies in the field
of non-Hermitian quantum physics, including that of exceptional points,
will  broaden our understanding of quantum mechanics.
Moreover, the results are expected to be of great value for applications.


\begin{thebibliography}{99}
\bibitem{local} In the present paper, the meaning of {\it localization} 
of the states is used  differently from that in \cite{shot,naz1}. 
Considering only the extension of the cavity, the whispering gallery modes are
localized (i.e. concentrated within a relatively small part) in contrast 
to the chaotic modes (which are distributed over the whole cavity).
However, the chaotic states are almost decoupled
from the continuum and therefore well localized {\it inside} 
the region  defined by the cavity, 
while the whispering gallery modes are strongly coupled to the continuum
and have non-negligible contributions from {\it outside} the cavity 
(i.e. from inside the attached leads).
\end{thebibliography}
\end{document}